\documentclass[12pt,preprint]{aastex}

\newcommand{\smallzero}{{\scriptscriptstyle 0}}
\newcommand{\smallone}{{\scriptscriptstyle 1}}
\newcommand{\smalltwo}{{\scriptscriptstyle 2}}
\newcommand{\smallA}{{\scriptscriptstyle A}}
\newcommand{\smallH}{{\scriptscriptstyle H}}
\newcommand{\smallM}{{\scriptscriptstyle M}}
\newcommand{\smallR}{{\scriptscriptstyle R}}
\newcommand{\smallphi}{{\scriptscriptstyle \phi}}

\newcommand{\vpttRz}{{\stackrel{\scriptscriptstyle \approx}{v}_\smallphi \!(R,z)}}
\newcommand{\xiRttRz}{{\stackrel{\scriptscriptstyle \approx}{\xi}_\smallR (R,z)}}

\newcommand{\zbb}{\bar{\bar{\mbox{$z$}}}}

\newcommand{\gam}{\gamma_\smallzero}
\newcommand{\gamb}{\bar{\gamma}_\smallzero}
\newcommand{\gamk}{\gamma_{\smallzero k}}
\newcommand{\po}{\rho_\smallzero}
\newcommand{\ko}{k_\smallzero}
\newcommand{\kR}{k_\smallR}
\newcommand{\Ho}{H_\smallzero}
\newcommand{\Ro}{R_\smallzero}
\newcommand{\kzo}{k_z^\smallzero}
\newcommand{\GamJ}{\Gamma_J}
\newcommand{\GamJk}{\Gamma_{J k}}
\newcommand{\GamJh}{\hat{\Gamma}_J}
\newcommand{\GamJt}{\tilde{\Gamma}_J}
\newcommand{\Deff}{\mathcal{D}_{\mathrm{eff}}}

\newcommand{\vaz}{v_{\smallA z}}
\newcommand{\vap}{v_{\smallA \smallphi}}
\newcommand{\waz}{\omega_{\smallA z}}
\newcommand{\aH}{\alpha_{\smallH}}
\newcommand{\va}{v_\smallA}
\newcommand{\vp}{v_\smallphi}
\newcommand{\vph}{\hat{v}_\smallphi}
\newcommand{\vpt}{\tilde{v}_\smallphi}
\newcommand{\vR}{v_\smallR}
\newcommand{\vRh}{\hat{v}_\smallR}
\newcommand{\vpkh}{\hat{v}_{\smallphi k}}
\newcommand{\vRkh}{\hat{v}_{\smallR k}}
\newcommand{\BR}{B_\smallR}
\newcommand{\BRh}{\hat{B}_\smallR}
\newcommand{\Bp}{B_\smallphi}
\newcommand{\Bph}{\hat{B}_\smallphi}
\newcommand{\Bpkh}{\hat{B}_{\smallphi k}}
\newcommand{\BRkh}{\hat{B}_{\smallR k}}

\newcommand{\xiR}{\hat{\xi}_\smallR}
\newcommand{\xiz}{\hat{\xi}_z}
\newcommand{\xip}{\hat{\xi}_\smallphi}
\newcommand{\xiRkh}{\hat{\xi}_{\smallR k}}
\newcommand{\xipkh}{\hat{\xi}_{\smallphi k}}
\newcommand{\xiRt}{\tilde{\xi}_\smallR}
\newcommand{\xizt}{\tilde{\xi}_z}
\newcommand{\xiRto}{\tilde{\xi}_{\smallR , \smallzero}}
\newcommand{\xiRtM}{\tilde{\xi}_{\smallR \smallM}}
\newcommand{\xizto}{\tilde{\xi}_{z,\smallzero}}
\newcommand{\xipt}{\tilde{\xi}_\smallphi}

\newcommand{\ddz}{\frac{\partial}{\partial z}}

\newcommand{\ddzz}{\frac{\partial^2}{\partial z^2}}

\newcommand{\ddzzz}{\frac{\partial^3}{\partial z^3}}

\newcommand{\ddR}{\frac{\partial}{\partial R}}
\newcommand{\ddRR}{\frac{\partial^2}{\partial R^2}}
\newcommand{\ddt}{\frac{\partial}{\partial t}}

\newcommand{\DDzb}{\frac{d}{d \bar{z}}}
\newcommand{\DDzz}{\frac{d^2}{d z^2}}
\newcommand{\DDzzb}{\frac{d^2}{d \bar{z}^2}}
\newcommand{\DDzzbb}{\frac{d^2}{d \zbb^2}}
\newcommand{\DDzzzzb}{\frac{d^4}{d \bar{z}^4}}

\begin{document}

\title{Ballooning Modes in Thin Accretion Disks: \\
Limits for their Excitation}

\author{B. Coppi and E.A. Keyes}
\affil{Massachusetts Institute of Technology, Cambridge, MA 02139}

\shorttitle{Ballooning Modes in Thin Accretion Disks}
\shortauthors{Coppi and Keyes}

\begin{abstract}

The conditions that limit the possible excitation of ideal MHD axisymmetric
ballooning modes in thin accretion disks are discussed.  As shown earlier
by \citet{coppia}, these modes are well-localized in the vertical
direction but have characteristic oscillatory and non-localized profiles in
the radial direction.  A necessary condition for their excitation is that
the magnetic energy be considerably lower than the thermal energy.  Even
when this is satisfied, there remains the problem of identifying the possible
physical factors which can make the considered modes radially localized.  The
general solution of the normal mode equation describing the modes is given,
showing that it is characterized by a discrete spectrum of eigensolutions.
The growth rates are reduced and have a different scaling relative to those 
of the ``long-cylinder'' modes, commonly known as the Magneto Rotational
Instability, that have been previously studied.

\end{abstract}

\keywords{MHD --- accretion, accretion disks --- magnetic fields --- methods: analytical}

\section{Introduction}
\label{sec:intro}

The problem of identifying the processes which can produce significant
transport of angular momentum outward in accretion disks has led to
consider the plasma collective modes that can be excited in rotating
plasmas where a magnetic field is present.  In these plasmas the angular
momentum is assumed to increase with the distance from the axis of
rotation as in the case of Keplerian accretion disks.  The most immediate
approach to this problem is that of considering modes that are axisymmetric
and are of the same type as those found originally for a long cylindrical
plasma \citep{velikhov,chandra,balbus}.

\citet{coppia,coppib} pointed out that these axisymmetric
modes when applied to a thin disk configuration become of the ballooning
type \citep{coppid} in the vertical direction, i.e.\ they cannot be
represented by a single Fourier harmonic, and acquire a characteristic
oscillation in the radial direction.  The radial wavelength is related to 
the distance over which the mode is localized vertically.  Thus the problem
of finding a localizing factor in the radial direction for the axisymmetric
modes was left unresolved.  As pointed out by \citet{coppic} the scaling
of the growth rates of these ballooning modes is different from that of
the original modes, which are appropriate for a long cylinder rather than
for a thin disk, and are reduced by the condition of vertical localization.

In the present paper we discuss the general solution of the two-dimensional
equation that describes the axisymmetric modes in thin plasma disks where
the magnetic energy density is much smaller than the thermal energy density.
Two categories of modes are identified: the localized ballooning modes that
have a discrete spectrum and the quasi-localized modes that have a continuous
spectrum of growth rates and related radial wavelenths.  These quasi-localized
modes can be considered to be physically significant (i.e.\ to be
sufficiently well-localized vertically) only for very large values of
$\beta^t_s \equiv c_s^2 / \va^2$ where $c_s$ is the sound velocity and
$\va$ the Alfv\'{e}n velocity.  As is well known, the ratio $H/R$ is
of the order of $c_s / \vp$ where $\vp = \Omega R$ and
$\Omega = \Omega(R)$ is the local rotation frequency.

When all the limitations are considered, even if a factor providing a
radial localization for this kind of mode has not been identified yet, the
rate of angular momentum transport that can be expected from its excitation
does not appear to have the magnitude of those semi-empirical forms of the
diffusion coefficients for the angular momentum transport that are used in
current models of disks.

In Section \ref{sec:deriv} the set of two-dimensional equations that give
the spatial profile of axisymmetric modes is derived pointing out all the
limits of the derivation.  In Section \ref{sec:equation} the vertical
profile of the mode is shown to be described by a fourth order differential
equation under realistic conditions.

In Section \ref{sec:solution}, we give different analytical derivations of
the localized solutions of this mode equation, corresponding to a discrete
spectrum of growth rates and of radial wave numbers, in order to better
describe the properties and the limitations of these solutions.  The
asymptotic decay of the modes in the vertical direction is proportional
to a Gaussian.

In Section \ref{sec:quasi} we introduce the class of ``quasi-localized''
modes whose asymptotic decay in the vertical direction is not proportional
to a Gaussian but is algebraic and that have a continuous spectrum.

In Section \ref{sec:obser} the issue of the mode radial localization is
discussed and the difficulty of constructing mode packets that could be
radially localized is pointed out.  We compare the ballooning modes to
non-normal mode packets which may be formed from MRI modes to achieve
vertical localization.  And finally the rate of angular momentum
transport that can be expected from these modes is estimated and compared
to that corresponding to currently adopted effective diffusion coefficients.

\section{Mode Profile Equation}
\label{sec:deriv}

We adopt cylindrical coordinates and note that axisymmetric modes are
represented by the perturbed toroidal velocity
\begin{displaymath}
\vph = \; \vpttRz \; \exp (\gam t)\;.
\end{displaymath}
In order to identify the properties of $\vpttRz$ we start by writing the
total momentum conservation equation as
\begin{equation}
\label{Am}
\hat{\mathbf{A}}_m \equiv \rho \left( \ddt
\hat{\mathbf{v}} + \mathbf{v} \cdot \nabla \hat{\mathbf{v}} +
\hat{\mathbf{v}} \cdot \nabla \mathbf{v} \right) + \nabla
\left( \hat{p} + \frac{1}{4 \pi} \mathbf{B} \cdot \hat{\mathbf{B}} \right)
- \frac{1}{4 \pi} \mathbf{B} \cdot \nabla  \hat{\mathbf{B}} \simeq 0
\end{equation}
where we neglect the $R$ derivatives of all equilibrium quantities when
compared to the $R$ derivatives of the perturbation.  Then, as is customary
in the theory of ideal MHD modes, we consider the
$\mathbf{e}_\phi \cdot \nabla \times \hat{\mathbf{A}}_m = 0$
component of Eq.\ (\ref{Am}).  This is
\begin{displaymath}
\ddz \hat{A}_{mR} - \ddR \hat{A}_{mz} = 0
\end{displaymath}
and, specifically,
\begin{equation}
\label{phicurl}
\ddz \left[ \rho \left( \gam \vRh -
2 \Omega \vph \right) - \frac{1}{4 \pi} B_z
\ddz \BRh \right] -
\ddR \left[ \gam \rho \hat{v}_z -
\frac{1}{4 \pi} B_z \ddz \hat{B}_z \right] = 0
\end{equation}

We note that, as shown by \citet{coppia,coppib}, $\vph$
and the plasma displacement $\xiRt$ do not vanish when
$\gam / \Omega \rightarrow 0$, while $\vRh = \gam \xiR$,
$\hat{v}_z = \gam \xiz$, and the compressibility
\begin{displaymath}
\nabla \cdot \mathbf{v} \simeq \gam \left( \ddR \xiR + \ddz \xiz \right)
\simeq \gam \nabla \cdot \hat{\xi}
\end{displaymath}
do.  For the modes of interest the variation of $\vpttRz$ on $z$ occurs on
a scale distance that is considerably shorter than that for the variation
of $B_z$.  This is consistent with the fact that, in the considered region,
the $\BR$ component of the field is negligible.

The frozen-in law can be written in the form
\begin{equation}
\label{frozenin}
\frac{\partial \hat{\mathbf{B}}}{\partial t} =
\left( \mathbf{B} \cdot \nabla \hat{\mathbf{v}} + \hat{\mathbf{B}}
\cdot \nabla \mathbf{v} \right) - \mathbf{B} \left( \nabla \cdot
\hat{\mathbf{v}} \right) - \left( \hat{\mathbf{v}} \cdot \nabla \mathbf{B} +
\mathbf{v} \cdot \nabla \hat{\mathbf{B}} \right)
\end{equation}
and the $R$ component of it yields
\begin{displaymath}
\BRh = B_z \ddz \xiR
\end{displaymath}
\begin{displaymath}
\ddz \hat{B}_z \simeq - \ddR \BRh \simeq - B_z \ddR \ddz \xiR \;.
\end{displaymath}
Thus Eq.\ (\ref{phicurl}) becomes
\begin{equation}
\label{phicurlprime}
\ddzz \left[ \rho \left( \gam^2 \xiR - 2 \Omega \vph \right)
- \frac{1}{4 \pi} B_z^2 \ddzz \xiR \right] -
\ddRR \left[ - \gam^2 \rho \xiR + \frac{1}{4 \pi} B_z^2 \ddzz \xiR \right]
- \gam^2 \rho \ddR \nabla \cdot \hat{\xi} \simeq 0 \;.
\end{equation}

Now, we write
\begin{displaymath}
\vph \equiv - \frac{d\Omega}{dR} R \xiR + \gam \xip
\end{displaymath}
considering the fact that for $\gam = 0$ the frozen-in law (\ref{frozenin})
gives $\vph = - (d\Omega / dR) R \xiR$.  Then the $\phi$ component
of Eq.\ (\ref{frozenin}) gives
\begin{displaymath}
\Bph = B_z \ddz \xip - \Bp \nabla \cdot \hat{\xi}
\end{displaymath}
and the $\phi$ component of Eq.\ (\ref{Am}) becomes
\begin{equation}
\label{momentumphi}
\rho \gam \left( \gam \xip + 2 \Omega \xiR \right) \simeq \frac{1}{4 \pi} B_z
\ddz \left( B_z \ddz \xip - \Bp \nabla \cdot \hat{\xi} \right) \;.
\end{equation}

Now the set of equations (\ref{phicurlprime}) and (\ref{momentumphi}) needs
to be completed by one that would relate $\nabla \cdot \hat{\xi}$ to
$\xiR$ and $\xip$.  Therefore we note that the adiabatic equation of state
gives
\begin{displaymath}
\hat{p} \simeq - p \Gamma \left( \nabla \cdot \hat{\xi} \right)
\end{displaymath}
and that by neglecting the $z$ derivatives of the equilibrium quantities
relative to those of the perturbations in the equation
$\partial \hat{A}_{mz} / \partial z = 0$ we obtain, for
$\partial \xiz / \partial z = \nabla \cdot \hat{\xi} - \partial \xiR /
\partial R$,
\begin{equation}
\label{Bdotprime}
\gam^2 \left( \nabla \cdot \hat{\xi} - \ddR \xiR \right) +
\ddzz \left[ \vap^2 \frac{\Bph}{\Bp} -
c_s^2 \left( \nabla \cdot \hat{\xi} \right) \right] \simeq 0
\end{equation}
with $\Gamma = 5/3$ and $c_s^2 \equiv \Gamma p / \rho$.

%
%
%

Since, as will be shown in the following analysis, the excitation of
axisymmetric modes is possible only if $\vap^2 \ll c_s^2$, we consider
regimes for which this is the case and
$\gam^2 \ll c_s^2 | \partial^2 / \partial z^2 |$.  Then
Eq.\ (\ref{Bdotprime}) reduces to
\begin{equation}
\label{deldotxi}
\ddzz \left( \nabla \cdot \hat{\xi} \right) \simeq
\frac{\gam^2}{c_s^2} \ddR \xiR + \frac{\vaz^2}{c_s^2} \frac{\Bp}{B_z}
\ddzzz \xip
\end{equation}
This indicates that the terms involving $\nabla \cdot \hat{\xi}$ can
be neglected in Eqs.\ (\ref{phicurlprime}) and (\ref{momentumphi}) and
the set of equations that describe the axisymmetric modes reduces to
\begin{equation}
\label{momentumphiprime}
\rho \gam \left( \gam \xip + 2 \Omega \xiR \right) =
\frac{1}{4 \pi} B_z^2 \ddzz \xip
\end{equation}
and
\begin{equation}
\label{phicurlprimeprime}
\ddzz \left[ \rho \left( \gam^2 \xiR - 2 \Omega \gam \xip + 2 \Omega
\Omega^\prime R \xiR \right) - \frac{1}{4 \pi} B_z^2 \left( \ddzz + \ddRR
\right) \xiR \right] + \ddRR \left( \gam^2 \rho \xiR \right) = 0 \;.
\end{equation}

In order to assess the influence of the modes that can be excited on the
effective rate of transport of angular moment in thin accretion disks, a
number of factors have to be considered.  Thus the most evident factors
to be taken into account in looking for the possible solutions of
Eqs.\ (\ref{momentumphiprime}) and (\ref{phicurlprimeprime}) are:
\newcounter{listcount}
\begin{list}{\roman{listcount})}{\usecounter{listcount}
\setlength{\leftmargin}{2\parindent}}
\item The mode growth rate
\item The vertical and radial characteristic scale distances
\item The value of the threshold for the mode onset
\item Whether or not a mode is contained (localized) vertically within the
disk and whether appropriate non-normal modes (packets) can be constructed
\item The physical features that may localize the mode radially.
\end{list}
In particular, it is well known that the rate of transport which can be
produced by excited plasma modes can be assessed in a rudimentary way from
an effective diffusion coefficient that includes both the mode growth rate
and its characteristic wavelenths.  Therefore modes with very short
wavelengths may be less effective that modes with longer wavelengths even
if the growth rates of the latter modes are smaller.

For the sake of simplicity, we do not consider at first the condition that
a mode should be vertically localized within a disk nor the criterion that
it should have the lowest threshold $\Omega_c$ possible for its onset
($\Omega > \Omega_c$).  In particular, we refer to modes that are propagating
in the vertical direction, have vertical wavelenths $\lambda_z = 2 \pi / k_z$
that are much smaller than the height of the disk, $2H$, and involve the
central value of the particle density $\po \equiv \rho (z=0)$.  We assume
also that the modes are propagating radially.  Therefore
\begin{displaymath}
\vpttRz \simeq \vpt \exp \left[ i k_z z + i \kR \left( R - \Ro \right) \right]
\end{displaymath}
where $R = \Ro$ is the radius at which all equilibrium quantities entering
Eqs.\ (\ref{momentumphiprime}) and (\ref{phicurlprimeprime}) are evaluated,
and $k_z H \gg 1$.  Thus Eq.\ (\ref{momentumphiprime}) becomes
\begin{displaymath}
\left( \gam^2 + \waz^2 \right) \xipt = - \gam 2 \Omega \xiRt
\end{displaymath}
where $\waz^2 \equiv k_z^2 \vaz^2$, $\vaz^2 \equiv B_z^2 / (4 \pi \po)$,
and Eq.\ (\ref{phicurlprimeprime})
\begin{displaymath}
\left( \gam^2 + \waz^2 \right) k^2 \xiRt - 3 \Omega^2 k_z^2 \xiRt -
2 \Omega \gam \xipt k_z^2 = 0
\end{displaymath}
where $k^2 \equiv \kR^2 + k_z^2$ and we have taken
$2 \Omega^{\prime} \Omega R = -3 \Omega^2$ as appropriate for a Keplerian
disk.  The resulting dispersion relation is
\begin{equation}
\label{MRIdispersion}
\gam^4 + \gam^2 \left( 2 \waz^2 + \Omega^2 \frac{k_z^2}{k^2} \right) =
\waz^2 \left( 3 \Omega^2 \frac{k_z^2}{k^2} - \waz^2 \right) \;.
\end{equation}
We see that the instability threshold is $3 \Omega ^2 = k^2 \vaz^2$,
implying very high values of $\beta_s \equiv c_s^2 / \vaz^2$ as
$k^2 > k_z^2 \gg 1 / H^2$, and that the maximum growth rate ($\gam$
relatively close to $\Omega$) is found for $\kR^2 \ll k_z^2$.  This is
the limit in which the Velikhov (a.k.a.\ MRI) instability is found.  The
question then arises whether the assumptions that have to be made to arrive
at Eq.\ (\ref{MRIdispersion}) can make these types of modes good candidates
to produce the needed rate of transport in thin accretion disks.  Therefore,
in the next section we look for the modes that can be contained in the disk
vertically and have the lowest threshold possible.

\section{Contained Modes}
\label{sec:equation}

Now we analyze the characteristics of the modes that are localized
near the center of the disk over a distance smaller than $H$ where the
density $\rho(z)$ can be approximated by
\begin{equation}
\label{density}
\rho \simeq \po \left( 1 - \frac{z^2}{\Ho^2} \right)
\end{equation}
and $H \sim \Ho$.

These modes, as was demonstrated by \citet{coppia,coppib}, are characterized
by being evanescent in the vertical direction and oscillatory in the radial
direction and are represented by
\begin{displaymath}
\xiRttRz \simeq \xiRt (z) \exp \left[ -i \kR \left( R - \Ro \right) \right]
\end{displaymath}
where $\xiRt (z)$ is an even or odd function of $z$ such that
$\xiRt (|z| \rightarrow H) \rightarrow 0$.  Two other important points are
that, contrary to the case where Velikhov modes with the highest growth
rates can be found,
\begin{displaymath}
\kR^2 > \left| \frac{1}{\xiRt} \ddzz \xiRt \right|
\end{displaymath}
and the relevant growth rates \citep{coppic} scale as
\begin{equation}
\label{gammasmall}
\gam^2 \sim \frac{\vaz^2}{\Ho^2} < \Omega^2 \;.
\end{equation}
Therefore axisymmetric modes that are contained within the disk have a
characteristic growth rate that depends on the scale height of the disk
and is well below that of the MRI instability.

We note that, as shown by Eq.\ (\ref{deldotxi}), $|\nabla \cdot \hat{\xi}|
\ll | \partial \xiRt / \partial R |$.  Therefore $i \kR \xiRt \simeq
-d \xizt / dz$,
\begin{equation}
\label{incompressibility}
\xizt (z) \simeq i \kR \int\limits_{-\infty}^z \xiRt (z^\prime) dz^\prime
\;,
\end{equation}
and $\xizt (z)$ has the opposite parity of $\xiRt (z)$.  Moreover, since
$\xizt (z)$ has to be localized, like $\xiRt (z)$, within the disk the
solution $\xiRt (z)$ is subject to the condition
\begin{equation}
\label{xiz_relation}
\int\limits_{-\infty}^{+\infty} \xiRt (z^\prime) dz^\prime = 0 \;.
\end{equation}

The problem to be solved is simplified if we note that 
Eq.\ (\ref{momentumphiprime}) reduces to
\begin{equation}
\label{xip_relation}
2 \Omega \gam \xiRt \simeq \vaz^2 \DDzz \xipt
\end{equation}
by observing that $\gam^2 \sim \vaz^2 / \Ho^2$ and that the modes we
look for are localized within $-\Ho < z < \Ho$:
\begin{displaymath}
\Ho^2 \left| \frac{1}{\xipt} \DDzz \xipt \right|^2 \gg 1 \;.
\end{displaymath}
Under the same conditions Eq.\ (\ref{phicurlprimeprime}) reduces to
\begin{displaymath}
\DDzz \left[ \left( \kR^2 \vaz^2 - 3 \Omega^2 + \gam^2 \right) \xiRt +
\frac{3 \Omega^2 z^2}{\Ho^2} \xiRt - 2 \gam \Omega \xipt -
\vaz^2 \DDzz \xiR \right] - \gam^2 \kR^2 \xiRt \simeq 0 \;.
\end{displaymath}
Thus the equation for the mode amplitude is reduced to one of the fourth
order, that is
\begin{displaymath}
\DDzz \left[ \left( \kR^2 \vaz^2 - 3 \Omega^2 + \gam^2 \right) \xiRt +
\frac{3 \Omega^2 z^2}{\Ho^2} \xiRt - \vaz^2 \DDzz \xiRt \right] 
- \left( \gam^2 \kR^2 + \frac{4 \gam^2 \Omega^2}{\vaz^2} \right)
\xiRt \simeq 0 \;.
\end{displaymath}
Referring to Eq.\ (\ref{xip_relation}) we see that $\xipt \neq 0$ only
when $\gam \neq 0$, and in this case the condition that 
$\xipt (|z| \rightarrow H) \rightarrow 0$ implies that
\begin{equation}
\label{doubleintegral}
\int\limits_{-\infty}^{+\infty} dz^{\prime}
\int\limits_{-\infty}^{z^{\prime}} \xiRt (z^{\prime \prime})
dz^{\prime \prime} = 0 \;.
\end{equation}
Therefore when $\gam \neq 0$, $\xiRt (z)$ has the two constraints
(\ref{xiz_relation}) and (\ref{doubleintegral}) in addition to that
of vertical containment $\xiRt (|z| \rightarrow H) \rightarrow 0$.

Now we define
\begin{displaymath}
\ko^2 \equiv \frac{3 \Omega^2}{\vaz^2} \quad , \quad
\Delta_z \equiv \sqrt{\frac{\Ho}{\ko}} \sim \Ho \sqrt{\frac{\vaz}{c_s}}
\end{displaymath}
with $\bar{z} \equiv z / \Delta_z$ and obtain
\begin{displaymath}
\DDzzb \left[ \DDzzb \xiRt - \bar{z}^2 \xiRt - \left(
\frac{\kR^2}{\ko^2} - 1 + \frac{\gam^2}{3 \Omega^2} \right)
\ko \Ho \xiRt \right] + \left( \frac{\kR^2}{\ko^2}
+ \frac{4}{3} \right) \frac{\gam^2 \Ho^2}{\vaz^2} \xiRt \simeq 0 \;.
\end{displaymath}

It is evident that the solution is localized over a smaller distance
than $\Ho$ if $\ko^2 \Ho^2 \gg 1$.  We note also that
$\kR^2 / \ko^2 -1 \sim 1 / (\ko \Ho) \sim
\vaz/c_s \ll 1$, and we consider the asymptotic limit (\ref{gammasmall}).
Therefore we introduce two additional dimensionless quantities involving
the radial wave number and the growth rate
\begin{displaymath}
\Lambda  \equiv  - \left( \frac{\kR^2}{\ko^2} - 1  \right)
\ko \Ho
\end{displaymath}
\begin{displaymath}
\gamb^2  \equiv \left( \frac{\kR^2}{\ko^2} + \frac{4}{3} \right)
\frac{\gam^2 \Ho^2}{\vaz^2} \;.
\end{displaymath}
For $\kR \simeq \ko + \delta \kR$,
$\Lambda \simeq - 2 \delta \kR \Ho$.  The implied dimensionless form
of the normal mode equation is thus
\begin{equation}
\label{modeprofile}
\DDzzb \left[ \left( \DDzzb - \bar{z}^2 + \Lambda \right) \xiRt (\bar{z})
\right] + \gamb^2 \xiRt (\bar{z}) = 0 \;.
\end{equation}

Finally we note that the adopted linearized approximation is valid to the
extent that
$| \xizt | < \Delta_z$, implying $| \xiRtM | \sqrt{\ko \Ho} <
\sqrt{ \Ho / \ko}$, that is
\begin{displaymath}
\left| \xiRtM \right| < \frac{1}{\ko} \sim \frac{\vaz \Ho}{c_s} \;,
\end{displaymath}
where $| \xiRtM | = \max | \xiRt |$.  We note also that the condition
$\ko^2 / \kR^2 - 1 \ll 1$ implies that $\Lambda < \ko \Ho$.

\section{Solutions of the Normal Mode Equation}
\label{sec:solution}

The simplest case to consider is that of marginal stability, that is
$\gamb^2 \rightarrow 0$.  Then
Eq.\ (\ref{modeprofile}) reduces to
\begin{displaymath}
\left[ \DDzzb - \bar{z}^2 + \Lambda \right] \xiRt (\bar{z}) = 0
\end{displaymath}
with the boundary condition that $\xiRt(\bar{z})$ vanishes at large
$\bar{z}$.  The solutions of this equation are given by
\begin{displaymath}
\xiRt(\bar{z}) = \xiRto H_n(\bar{z}) \exp \left( - \frac{\bar{z}^2}{2} \right)
\end{displaymath}
for integer $n \ge 0$, where $H_n$ are the Hermite polynomials, with the
corresponding eigenvalues $\Lambda = 2n+1$.

We note that in order to comply with the condition that $\xizt$ be a localized
function of $z$, represented by Eq.\ (\ref{xiz_relation}), only odd-parity
$\xiRt (\bar{z})$ solutions are acceptable.  Thus the lowest eigenfunction
is represented by
\begin{displaymath}
\xiRt (\bar{z}) = \xiRto \bar{z} \exp \left( - \frac{1}{2} \bar{z}^2 \right)
\end{displaymath}
and
\begin{displaymath}
\xizt (\bar{z}) = \left( -i \left( \ko \Ho \right) ^{1/2} \xiRto
\right) \exp \left( - \frac{1}{2} \bar{z}^2 \right) \;.
\end{displaymath}
The corresponding eigenvalue is then $\Lambda = 3$ and
$\kR^2 = \ko^2 \left[ 1 - 3 / (\ko \Ho) \right]$.

\begin{figure}
\epsscale{0.5}
\plotone{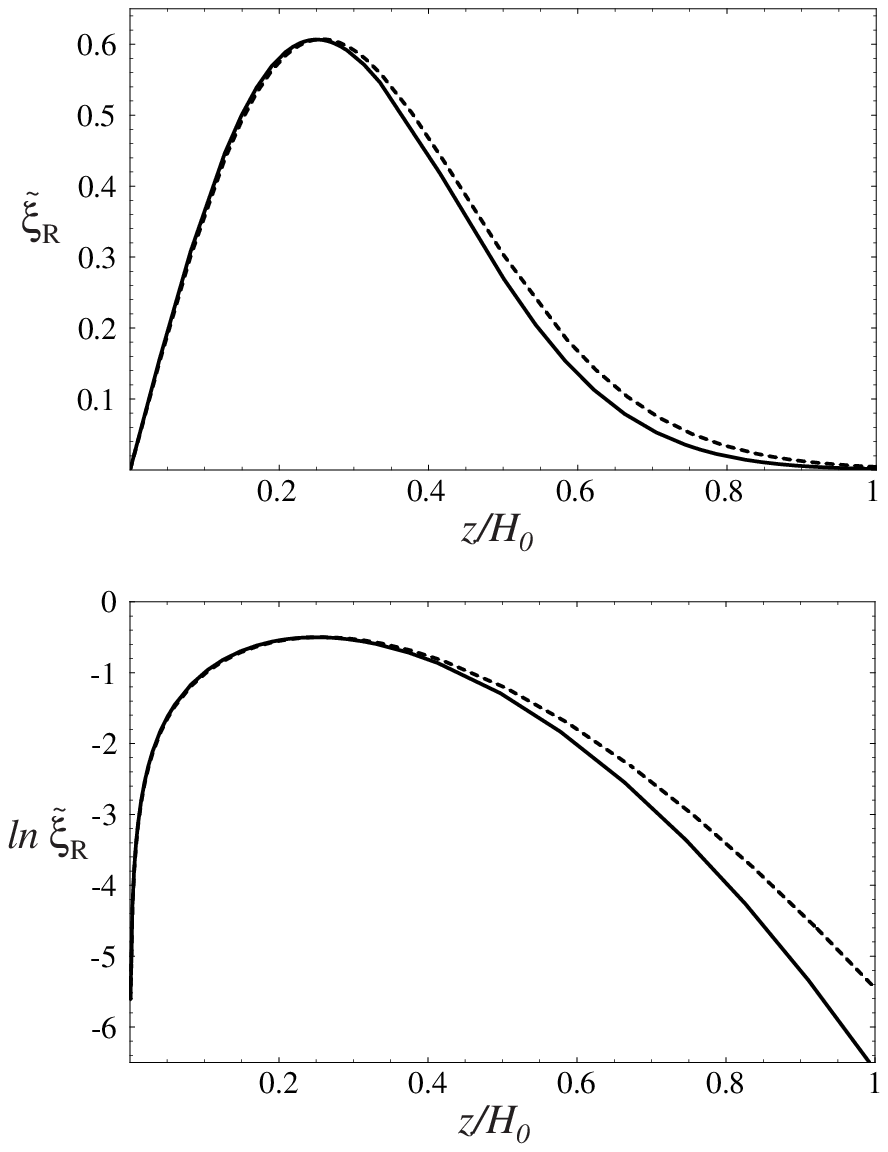}
\caption{
Corrections to the marginally stable solution.  The solid
line is the analytic marginally-stable solution for the approximation
(\ref{density}) of the density profile, giving $\Lambda=3$.  The
overlapping dotted line is a numerical solution for the mode in the case
of the density distribution $\rho = \po \exp (-z^2 / \Ho^2)$ that gives
$\Lambda \simeq 2.9$.  Here $\sqrt{\ko \Ho} = 4$, a relatively low value,
is chosen to emphasize the difference between the two solutions.
\label{fig:numerical}}
\end{figure}

In order to analyze a case where the solution extends beyond the values
of $z^2 / \Ho^2$ for which the approximation (\ref{density}) for the
particle density is valid, we consider the complete profile
$\rho = \po \exp (-z^2 / \Ho^2)$.  Then the equation for the
marginally-stable modes is
\begin{equation}
\label{gaussianden}
\left[ \exp \left( - \frac{z^2}{\Ho^2} \right) \right] \ko^2
\xiRt (z) - \kR^2 \xiRt (z) + \DDzz \xiRt (z) = 0
\end{equation}
and we see that $\kR$ cannot be too small relative to $\ko$ in order to
have the needed turning points within the main body of the disk.  Moreover,
if the solution is to be well-localized within the disk the condition
$(\ko \Ho)^2 \gg 1$ corresponding to $\beta_s \gg 1$ is necessary.  An
example of this is illustrated in Fig.\ \ref{fig:numerical} where
$\ko \Ho = 16$.  To illustrate this it may be useful to rewrite
Eq.\ (\ref{gaussianden}) as
\begin{displaymath}
\left[ \exp (-\zbb^2) - \frac{\kR^2}{\ko^2} \right]
\xiRt (\zbb) + \frac{1}{(\ko \Ho)^2} \DDzzbb \xiRt (\zbb) = 0 \;,
\end{displaymath}
where $\zbb \equiv z / \Ho$.  The solution given in Fig.\ \ref{fig:numerical}
has been obtained numerically and compared to the one that would be found
analytically replacing $\exp (-\zbb^2)$ by $1-\zbb^2$ for an equal value
of $\ko \Ho$.  \placefigure{fig:numerical}


Next we consider the case of unstable ($\gam \neq 0$) modes that can be
found for $\Lambda > 3$.  We note that the relevant solutions of
Eq.\ (\ref{modeprofile}) are subject to the conditions that $\xipt$ in
addition to $\xizt$ and $\xiRt$ vanish at large $\bar{z}$.
Therefore it is convenient to recast the equation in terms of the variable
$\tilde{Y} \propto \xipt$, considering that
$\tilde{Y}^{\prime} \propto \xizt$ and
$\tilde{Y}^{\prime \prime} \propto \xiRt$ given Eqs.\ (\ref{incompressibility})
and (\ref{xip_relation}).  Then we rewrite Eq.\ (\ref{modeprofile}) as a
set of two second-order differential equations
\begin{equation}
\label{firstdiff}
\DDzzb \tilde{Y} = \gamb \xiRt
\end{equation}
\begin{equation}
\label{seconddiff}
- \DDzzb \xiRt + \left( \bar{z}^2 - \Lambda \right) \xiRt = \gamb \tilde{Y}
\end{equation}
and note that this set of equations corresponds to the following matricial
equation
\begin{displaymath}
\renewcommand{\arraystretch}{0.75}
\left( \begin{array}{cc} O_{11} & 0 \\ 0 & O_{22} \end{array} \right)
\left( \begin{array}{c} \xiRt \\ \tilde{Y} \end{array} \right) =
\left( \begin{array}{cc} \Lambda & \gamb \\ \gamb & 0 \end{array} \right)
\left( \begin{array}{c} \xiRt \\ \tilde{Y} \end{array} \right)
\end{displaymath}
where the operators $O_{11}$ and $O_{22}$ are easily identifiable from
Eqs.\ (\ref{firstdiff}) and (\ref{seconddiff}).
This has a discrete spectrum of the eigenvalue matrix, when the
conditions that both $\xiRt$ and $\tilde{Y}$ are of a given parity and that
$\xiRt ( \bar{z}^2 \rightarrow \infty ) \rightarrow 0$ and 
$\tilde{Y} ( \bar{z}^2 \rightarrow \infty ) \rightarrow 0$.

Then we refer to the equation for $\tilde{Y}$
\begin{equation}
\label{diffeq}
\DDzzzzb \tilde{Y} - \left( \bar{z}^2 - \Lambda \right) \DDzzb \tilde{Y} +
\gamb^2 \tilde{Y} = 0
\end{equation}
that lends itself to derive the following quadratic form for $\xizt$
\begin{equation}
\label{quadratic}
\left< \left| \xizt^{\prime \prime} \right| ^2 + \bar{z}^2
\left| \xizt^{\prime} \right| ^2 \right> = \Lambda
\left< \left| \xizt^{\prime} \right| ^2 \right> + \gamb^2
\left< \left| \xizt \right| ^2 \right> \;.
\end{equation}
This is obtained multiplying Eq.\ (\ref{diffeq}) by 
$d^2 \tilde{Y} / d \bar{z}^2$ and integrating over 
$\left< - \infty , + \infty \right>$.

The localized solutions for $\bar{z}$ large are given, as in the case
where $\gamb \rightarrow 0$, by
\begin{displaymath}
\DDzzzzb \tilde{Y} - \bar{z}^2 \DDzzb \tilde{Y} \simeq 0 \;.
\end{displaymath}
Thus we have six constraints for the solution of Eq.\ (\ref{diffeq}):
normalization, two parity conditions, and three unwanted asymptotic limits,
for large $\bar{z}$, out of the expected four.  Six constraints lead to
two eigenvalue conditions: only specific values for both $\Lambda$ and
$\bar{\gamma}_0$ are allowed, as anticipated on the basis of
Eq.\ (\ref{quadratic}).

\begin{figure}
\epsscale{0.77}
\plotone{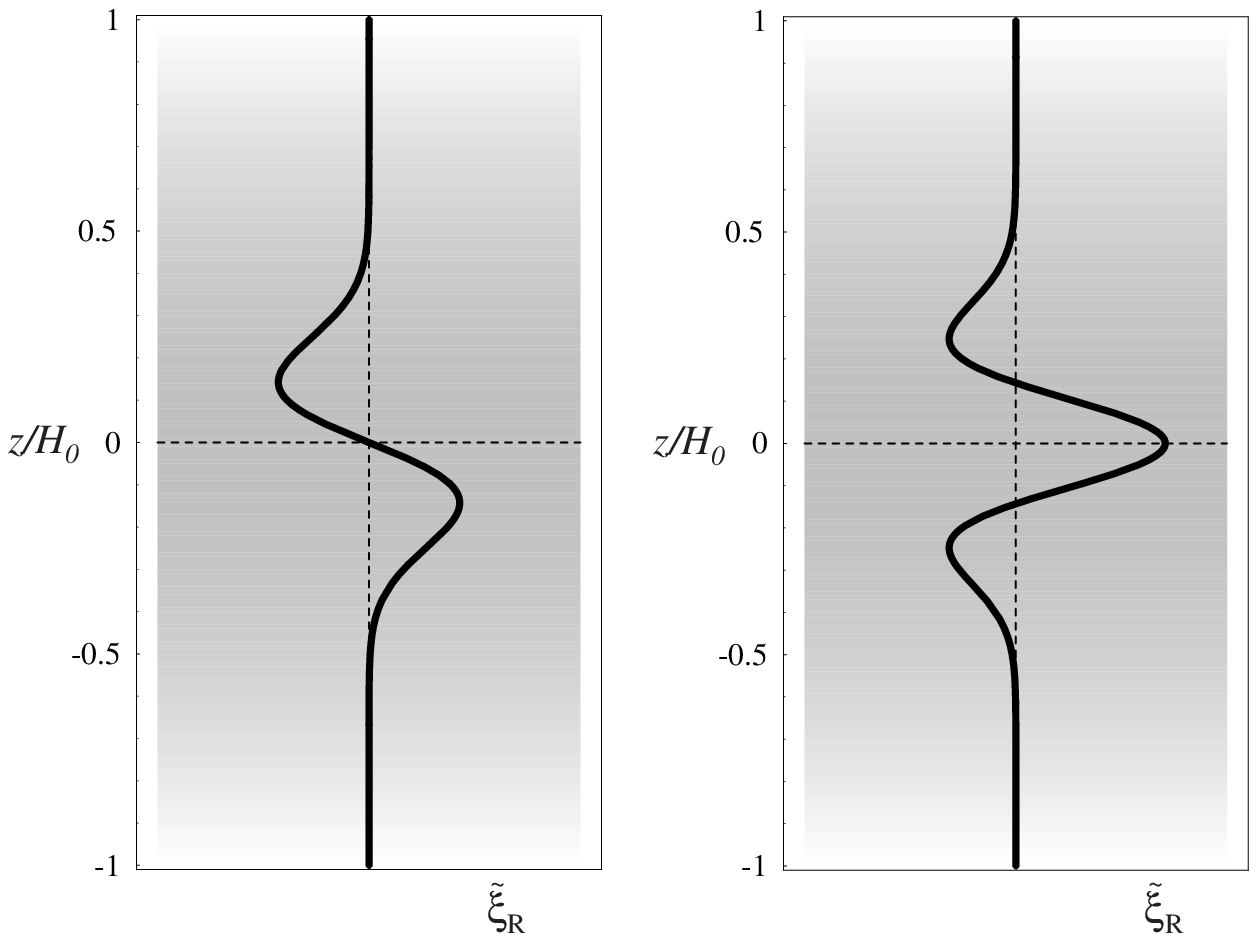}
\caption{
Vertical profiles of the marginally-stable mode (left)
and of the lowest unstable mode (right) for $\sqrt{\ko \Ho} = 7$.
Here $\ko = \sqrt{3} \Omega / \vaz$, and $\Ho$ is the
characteristic scale distance for the density vertical profile near
$z = 0$.  In the figure $z$ is scaled to $\Ho$.  The equilibrium
density profile is shown as a grayscale background.  The value of 
$\ko \Ho$ should be sufficiently large that the mode is
well-localized over $0 \le |z| \le \Ho$, {\em e.g.}
$\tilde{\xi} (z=\Ho) / \tilde{\xi} (z=0) \ll 0.1$.  Note that
the mode is localized over the width 
$\Delta_z \sim \Ho / \sqrt{\ko \Ho}$ and that
$(\ko \Ho)^{1/2} \sim \beta_s^{1/4}$.  Therefore the corresponding
values of $\beta_s$ are very large.
\label{fig:vertical}}
\end{figure}

\begin{figure}
\epsscale{0.77}
\plotone{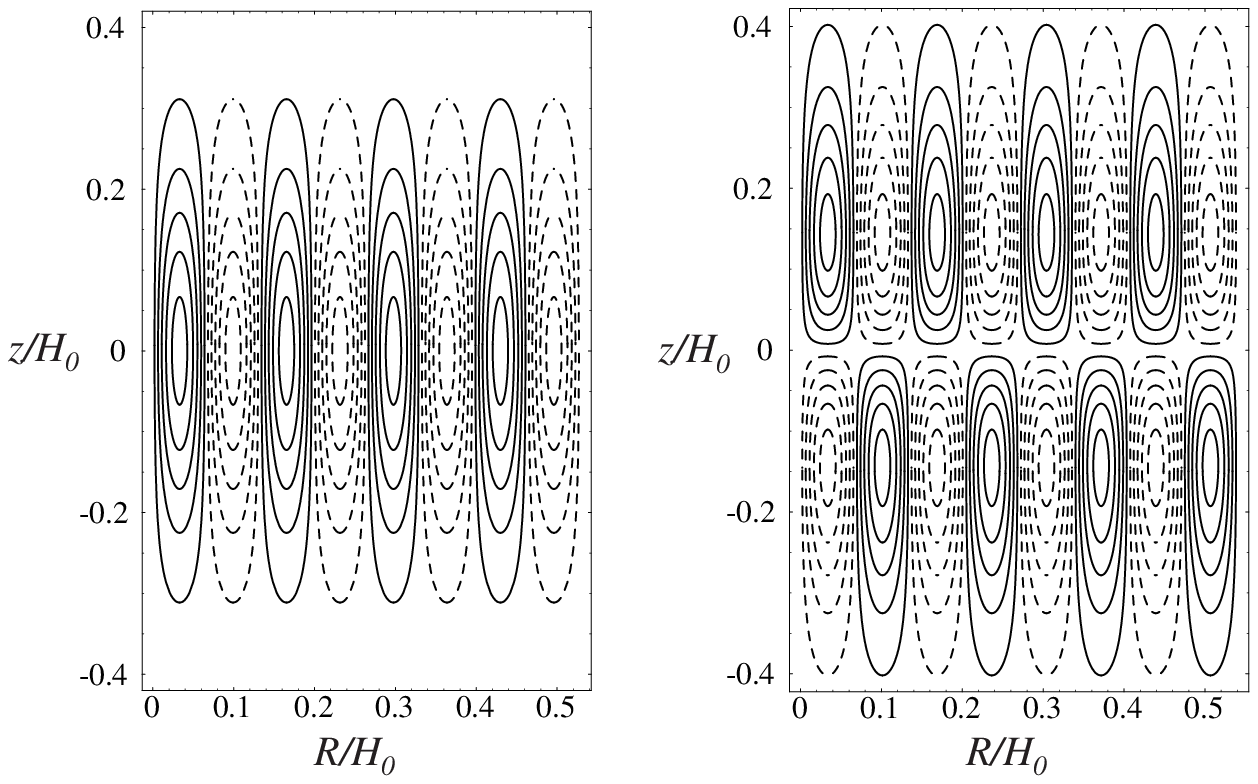}
\caption{
Two-dimensional streamline patterns of the plasma
displacement for the marginally-stable mode (left) and the lowest unstable
mode (right).  Solid versus dotted lines denote the sign of rotation.
Length units in the figure are scaled to $\Ho$.  As noted in the
text, the radial variation of the mode is faster than its vertical
variation, which is in turn stronger than the vertical variation of
the density profile.  Here $\sqrt{\ko \Ho} = 7$.
\label{fig:streamline}}
\end{figure}

The exact solutions that we have identified for Eq.\ (\ref{diffeq}) confirm
the general conditions and characteristics that we have outlined.  In
particular the lowest eigensolution is
\begin{equation}
\label{lowestgaussian}
\tilde{Y} = \exp \left( - \frac{\bar{z}^2}{2} \right) \propto \xipt
\end{equation} 
and corresponds to
\begin{displaymath}
\xizt = \xizto \bar{z} \exp \left( - \frac{\bar{z}^2}{2} \right)
\propto \frac{d \tilde{Y}}{d \bar{z}} \;,
\end{displaymath}
and
\begin{displaymath}
\xiRt = \xiRto \left( 1 - \bar{z}^2 \right) \exp \left( - \frac{\bar{z}^2}{2}
\right) \;,
\end{displaymath}
as shown in Figs.\ \ref{fig:vertical} and \ref{fig:streamline}.
\placefigure{fig:vertical} \placefigure{fig:streamline}  We note that,
for $\tilde{Y} (\bar{z})$ given by Eq.\ (\ref{lowestgaussian}),
$\tilde{Y}^{\prime \prime \prime \prime} = ( \bar{z}^2 - 1)
\tilde{Y}^{\prime \prime} + 2 \tilde{Y} + 4 \bar{z} \tilde{Y}^\prime$
and find $\Lambda = 5$, $\gamb^2 = 2$ from the condition
$3 - \bar{\gamma}_0^2 - \Lambda - (5-\Lambda) \bar{z}^2 = 0$.
Thus the relevant growth rate is
\begin{equation}
\label{lowestunstablegamma}
\gam = \left( \frac{6}{7} \right) ^ {1/2} \frac{\vaz}{\Ho} \ll \Omega
\end{equation}
and
\begin{equation}
\delta \kR = - \frac{5}{2 \Ho}
\end{equation}

The next eigenfunction $\tilde{Y} = \bar{z} \exp ( -\bar{z}^2 /2)$ corresponds
to $\xizt = \xizto (1-\bar{z}^2) \exp ( - \bar{z}^2/2)$ and 
$\xiRt = \xiRto \bar{z} (3-\bar{z}^2) \exp ( - \bar{z}^2/2)$.  The relevant
set of eigenvalues is $\Lambda = 7$, $\gamb^2 = 6$.  Moreover, there are two
even-parity solutions at $\Lambda = 9$, two odd solutions at $\Lambda = 11$,
three at 13 and 15, and so on up to higher and higher eigenvalues, as
indicated in Table \ref{modetable}.  \placetable{modetable}

We observe that the eigenfunctions with the highest values of $\gamb^2$
corresponding to the same value of $\Lambda$ (that is, the most unstable
solutions) are, in fact, given by
\begin{equation}
\label{hermite}
\tilde{Y} (\bar{z}) = H_n \left( \frac{\bar{z}}{\sqrt{2}} \right)
e^{-\bar{z}^2/2}
\end{equation}
\begin{equation}
\label{lambdan}
\Lambda = 5 + 2n
\end{equation}
\begin{equation}
\label{gamman}
\gamb^2 = (n+1)(n+2) \;.
\end{equation}

The problem can be dealt with more simply by noting that the Fourier
transform of the relevant solutions satisfy the following second-order
equation
\begin{displaymath}
k_{\bar{z}}^2 \frac{d^2}{dk_{\bar{z}}^2} \tilde{Y}(k_{\bar{z}}) +
4k_{\bar{z}}\frac{d}{dk_{\bar{z}}} \tilde{Y}(k_{\bar{z}}) +
(\Lambda k_{\bar{z}}^2 - \bar{\gamma}_0^2 + 2 - k_{\bar{z}}^4)\,\tilde{Y} = 0
\end{displaymath}
whose solutions corresponding to Eq.\ (\ref{hermite}) are
\begin{displaymath}
\tilde{Y}(k_{\bar{z}}) = k_{\bar{z}}^n \exp \left( -\frac{k_{\bar{z}}^2}{2}
\right)
\end{displaymath}
For these solutions we have
\begin{displaymath}
2+3n+n^2-\bar{\gamma}_0^2+(\Lambda-5-2n)\,k_{\bar{z}}^2 = 0
\end{displaymath}
which gives the conditions (\ref{lambdan}) and (\ref{gamman}).  Combining 
the two, we obtain the dispersion relation
\begin{equation}
\label{dispersion}
\gamb^2 = \frac{1}{4} ( \Lambda - 3) (\Lambda -1)
\end{equation}
where $\Lambda \geq 3$ takes integer odd values.

\section{Quasi-localized Modes}
\label{sec:quasi}

\begin{figure}
\epsscale{0.5}
\plotone{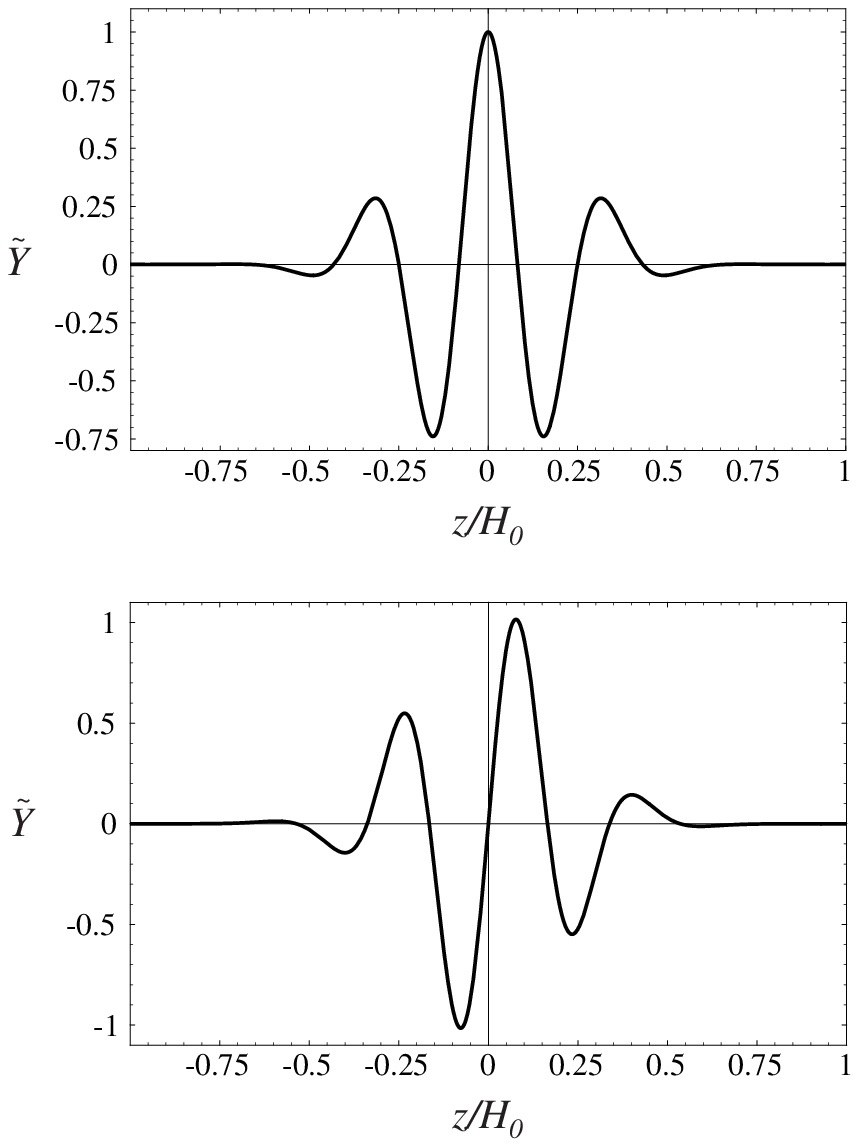}
\caption{Profiles of a localized even mode (upper) and of an odd
quasi-localized mode (lower) for $\Lambda = 13$ and $\gam^2 = 30$.  The
decay of the even mode is proportional to a Gaussian while that of the
odd mode is proportional to $\zbb^{-5}$, where $\zbb \equiv z / H$.
Here $\sqrt{\ko \Ho} = 7$.
\label{fig:quasi}}
\end{figure}

\begin{figure}
\epsscale{0.77}
\plotone{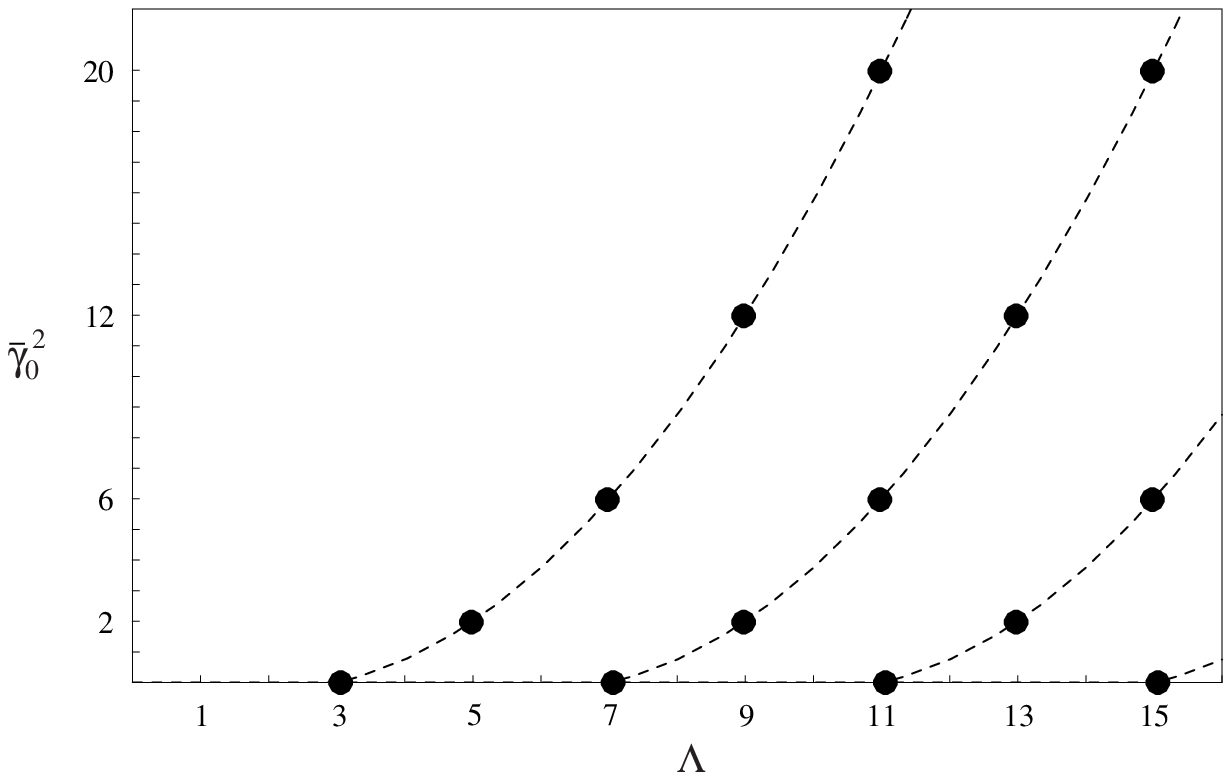}
\caption{
Graphic representation of the dispersion relation for the
axisymmetric ballooning modes.  The marginally-stable modes ($\gamb = 0$)
and the unstable modes ($\gamb > 0$) corresponding to discrete and odd
values of $\Lambda$ are indicated by dots.  Additional branches of
solutions with similar behavior exist at multiples of 4 at higher $\Lambda$.
Clearly the highest branch corresponding to the highest growth rates for
a given value of $\Lambda$ is the meaningful one.  The continuous spectrum
of quasi-localized modes is indicated by a dotted line, but in practice
they cannot be considered as physically acceptable unless $\gamb^2 \geq 30$,
a value that exceeds the asymptotic conditions under which the mode
equation (\ref{diffeq}) has been derived.
\label{fig:dispersion}}
\end{figure}

In addition to the well-localized modes with the discrete spectrum discussed
in the previous section, there is a class of modes to consider which decay
algebraically rather than exponentially as a function of $\bar{z}$ and that
have a continuous spectrum (see Figs. \ref{fig:quasi} and
\ref{fig:dispersion}).  Therefore we may classify them as quasi-localized
modes and, in practice, consider only those which have relatively fast
power-law decays such as $\tilde{Y} \sim 1/{\bar{z}}^\alpha$ with
$\alpha > 5$ as physically relevant.

The asymptotic solutions for $|\bar{z}| \rightarrow \infty$ of the mode
equation that depend algebraically on $\bar{z}$ are given by
\begin{displaymath}
\DDzzb \tilde{Y} - \bar{\gamma}_0^2 \tilde{Y} \simeq 0 \;,
\end{displaymath}
and are of the form
\begin{equation}
\label{power_law}
\tilde{Y} \propto \bar{z}^{1/2 \pm \sqrt{1/4 + \bar{\gamma}_0^2}}
\end{equation}
The ``practical'' condition $\alpha>5$ corresponds to $\gamb^2>30$ which
can be considered to exceed the asymptotic limit (\ref{gammasmall}) for
which the mode equation (\ref{diffeq}) has been derived, unless
$\ko^2 \Ho^2$ has extremely large values such that
$30 < \gamb^2 \ll \ko \Ho$.

The four asymptotic solutions of the fourth-order equation
are: decaying exponential, decaying power-law, growing exponential, and
growing power-law.  The last two are of course unacceptable.  Excluding
only the growing solutions, we are left with five conditions
on a fourth-order equation.  This leads to one eigenvalue condition,
corresponding to $\bar{\gamma}_0$ being a continuous function of $\Lambda$.

Since the discrete solutions are a special case of the continuous
spectrum formed by a linear combination of Gaussian-decaying and
power-law decaying asymptotic forms, we consider that
Eq.\ (\ref{dispersion}) remains valid for continuously varying
values of $\Lambda$.  Then we note that $\gamb^2>30$ corresponds to
$\Lambda>13$.

We may observe that when $\Lambda$ and $\gamb^2$ are related
by Eq.\ (\ref{dispersion}), Eq.\ (\ref{diffeq}) can be factored as follows:
\begin{equation}
\label{factoreddiffeq}
\left( \DDzzb - \bar{z} \DDzb + \frac{\Lambda -1}{2} \right)
\left( \DDzzb \tilde{Y} + \bar{z} \DDzb \tilde{Y} + \frac{\Lambda -3}{2}
\tilde{Y} \right) = 0
\end{equation}
This factorization separates the growing from the decaying
asymptotic solutions.  Thus
\begin{displaymath}
\tilde{Y} \simeq c_\smallone \exp \left( - \frac{\bar{z}^2}{2} \right) +
c_\smalltwo \frac{1}{\bar{z}^{(\Lambda - 3)/2}}
\end{displaymath}
The discrete spectrum corresponds to values of $\Lambda$ for which
$c_\smalltwo = 0$.  A relevant case is illustrated in Fig.\ \ref{fig:quasi}.
\placefigure{fig:quasi}

We note that the second-order equation resulting from
Eq.\ (\ref{factoreddiffeq}) can be solved analytically in general.
Introducing the variable $u \equiv - \bar{z}^2 /2$, we have
\begin{displaymath}
u \frac{d^2}{d u^2} \tilde{Y}(u) + \left( \frac{1}{2} - u \right)
\frac{d}{du} \tilde{Y}(u) - \left( \frac{\Lambda-3}{4} \right) \tilde{Y}(u) =0
\end{displaymath}
which is the standard differential equation for the confluent hypergeometric
function ${}_1F_1$, also known as the Kummer function:
\begin{displaymath}
\tilde{Y}_{\mathrm{even}}(\bar{z}) = {}_1F_1 \left( 
\frac{\Lambda -3}{4} , \frac{1}{2} , - \frac{\bar{z}^2}{2} \right)
\end{displaymath}
\begin{displaymath}
\tilde{Y}_{\mathrm{odd}}(\bar{z}) = \bar{z}\; {}_1F_1 \left(
\frac{\Lambda -1}{4} , \frac{3}{2} , - \frac{\bar{z}^2}{2} \right)
\end{displaymath}
For discrete choices of $\Lambda$ and particular parities, these Kummer
functions simplify to the localized solutions of Eq.\ (\ref{hermite}), but
in general they have power-law decays at large $\bar{z}$, and solutions
exist of either parity at the same eigenvalues.  \placefigure{fig:dispersion}

\section{Observations}
\label{sec:obser}



For the modes that have been presented, it is clear that $\Lambda$ will have 
to be limited by the condition that 
$\gam^2 \sim \vaz^2 / \Ho^2 \ll \Omega^2$.  Therefore unless 
$\vaz^2 / c_s^2$ is exceedingly small very few modes can be fitted in the
interval $5 \le \Lambda \ll \ko \Ho \sim \beta_s^{1/2}$.  Thus a problem which
arises is that of identifying a physical factor which can localize radially
the normal modes considered; a mode packet, constructed by varying $\Lambda$
over a significant interval, is difficult to envision.

We note also that, by comparison with the ``long cylinder'' modes, the
``cost'' of localizing a mode vertically between $z = -H$ and $z = +H$
and creating the needed turning points in the appropriate equation
involves (i) the requirement that $\kR$ be significant and related to
the scale of vertical localization $\Delta_z \sim (\Ho / \ko
)^{1/2}$ with $\kR \Delta_z \gg 1$ and $\kR \simeq \ko$ ;
(ii) a considerable reduction of the growth rate $\gam$ relative to
$\Omega$ and its dependence on the scale distance $\Ho$.  If a physical
factor providing the radial localization of the discrete modes given by
Eq.\ (\ref{diffeq}) can be found it is reasonable to expect that the resulting
growth rate will depend on a second scale distance and that it should be
depressed further relative to the expression (\ref{lowestunstablegamma}).
In order to resolve the issue of the radial localization of the mode, the best
option may be that of attempting to superpose modes centered around a range
of values of $\Ro$, taking into account that the corresponding values
of $\kR$, $\gam$, and $\Delta_z$ vary as a function of $\Ro$.



\begin{figure}
\epsscale{0.77}
\plotone{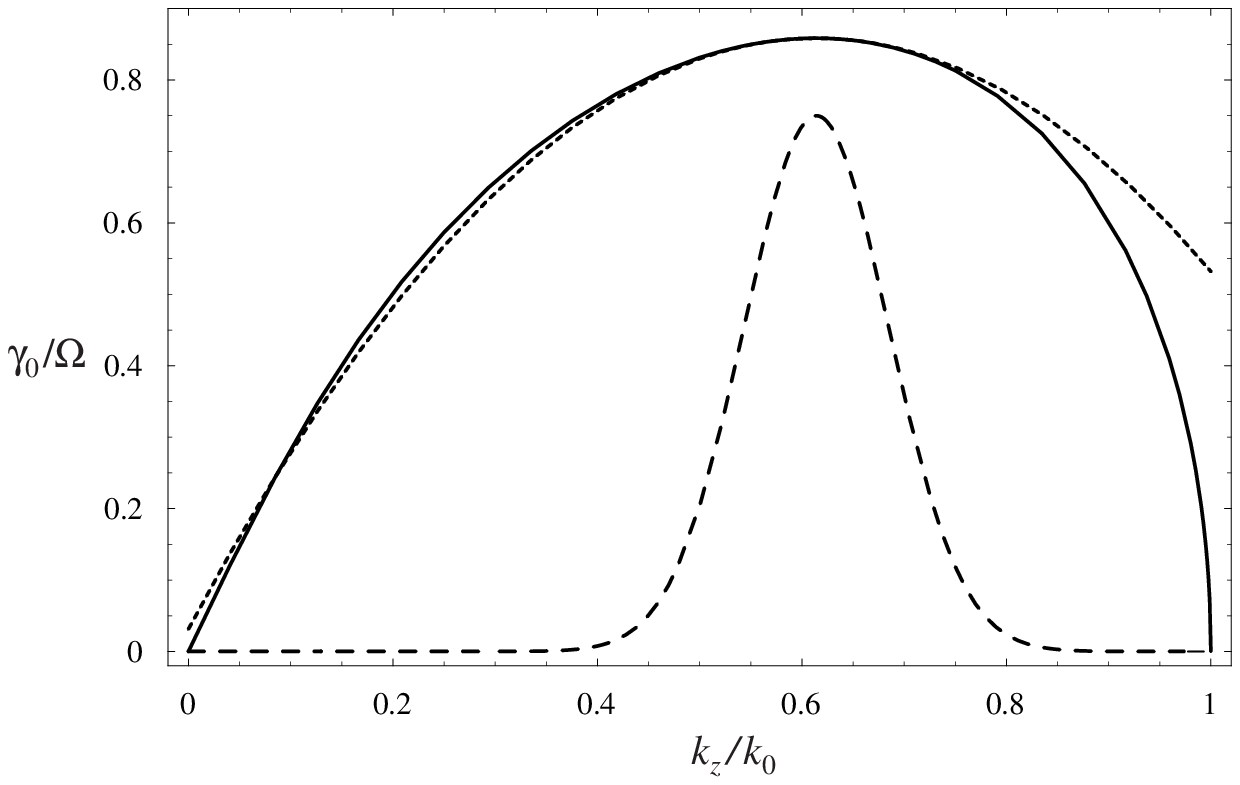}
\caption{
Dispersion relation of the MRI modes.  This plot shows the growth rate
of the MRI modes as a function of $k_z$ when $\kR=0$.  The maximum growth
rate of $\gam \simeq 0.85 \Omega$ occurs at $k_z \simeq 0.6 \ko$, and
the region of instability is bounded by $\kR^2 + k_z^2 < \ko^2$.  Here
$\ko = \sqrt{3} \Omega / \vaz$. The
short dashed line gives the approximation (\ref{disp_expansion}) to the
dispersion relation.  The long dashed line is a sample wavepacket with
$\Delta = 10 \ko^{-1}$ centered around the peak growth rate, shown for
comparison with the accuracy of the approximation.  Clearly, narrower
spectra for $k_z$ could be chosen, but they would correspond to larger
values of $\Delta$.
\label{fig:MRIdispersion}}
\end{figure}

\placefigure{fig:MRIdispersion}

The argument could be made that by taking very small wavelengths, requiring
that very large values of $\beta_s$ be considered, the MRI dispersion
relation (\ref{MRIdispersion}) could be used with $\kR^2 \ll k_z^2$ (see
Fig.\ \ref{fig:MRIdispersion}).  Then
``large'' growth rates would be obtained, but wave packets would have to
be constructed to ensure that the resulting non-normal mode would be
localized vertically.  However, we can argue easily that a packet can spread
rapidly, losing its initial localization property.  In particular we consider
an interval of values of $k_z$ around $k_z = \kzo$ where the growth rate
given by Eq.\ (\ref{MRIdispersion}) is maximum.  Therefore in this interval,
\begin{equation}
\label{disp_expansion}
\gam (k_z) \simeq \gam (\kzo) + \frac{1}{2}
\frac{\partial^2 \gam}{\partial k_z^2} (k_z - \kzo)^2 \; ,
\end{equation}
where $\partial^2 \gam / \partial k_z^2 < 0$, and
$\partial \gam / \partial k_z = 0$ for $k_z = \kzo$.  Then we construct the
packet
\begin{eqnarray}
\Psi & \propto & \left[ \int d k_z \exp \left\{ - \Delta^2 (k_z - \kzo)^2 +
\frac{1}{2} \frac{\partial^2 \gam}{\partial k_z^2} t (k_z - \kzo)^2 +
i (k_z - \kzo) z \right\} \right] \\ \nonumber
& & \exp \Big[ \gam (\kzo) t + i \kzo z + i \kR (R - \Ro) \Big]
\end{eqnarray}
and we obtain
\begin{equation}
\label{modepacket}
\Psi \propto \exp \left\{ -\frac{z^2}{4 \left[ \Delta^2 - 
\frac{1}{2} \frac{\partial^2 \gam}{\partial k_z^2} t \right]} +
\gam t + i \kzo z \right\}
\end{equation}
where $\Delta^2 \ll H^2$.  It is evident that the packet begins to spread
for ${\kzo}^2 (\partial^2 \gam / \partial k_z^2) t \sim 2 \Delta^2
{\kzo}^2$.

In order to estimate the rate of angular momentum transport that these
packets may produce we derive the flux of angular momentum from the
quasilinear theory of the unconfined normal modes that are the mode packet
components.  Clearly,
\begin{displaymath}
\GamJ = R \left[ \rho \vp \vR - \frac{1}{4 \pi} \Bp \BR \right] \; .
\end{displaymath}

Considering the perturbations from the equilibrium state, the flux $\GamJ$
can be separated into an average and a fluctuating part:
$\GamJ = \left< \GamJ \right> + \GamJh$ where
$\big< \GamJt \big> = 0$.  In particular
$\big< \GamJ \big> = \sum_k \big< \GamJk \big>$ and we note that
the particle flux associated with these modes $\big< \Gamma_p \big>$ is
null as $\big< \hat{\rho} \vRh \big> = 0$ given the fact that
$\nabla \cdot \mathbf{\hat{v}} = 0$ and $\hat{\rho} = 0$.  Then
\begin{equation}
\left< \GamJk \right> = \rho \left< \vpkh^{} \vRkh^{*} \right> -
\frac{1}{4} \left< \Bpkh^{} \BRkh^{*} \right> + \mathrm{c.c.}
\end{equation}
where $\vRkh = \gamk \xiRkh$,
$\vpkh = - \Omega^{\prime} R \xiRkh + \gamk \xipkh$,
$\BRkh = i (k_z B_z) \xiRkh$, and $\Bpkh = i (k_z B_z) \xipkh$.
For $G_k \equiv \big< \GamJk / (\rho R) \big>$ we obtain
\begin{equation}
G_k = 2 \gamk \left< \left| \xiRkh \right|^2 \right> \left[ -\Omega^{\prime} R
+ 2 \Omega \frac{\waz^2 - \gamk^2}{\waz^2 + \gamk^2} \right]
\end{equation}
and the relevant effective diffusion coefficient can be defined as
\begin{equation}
\Deff \simeq 2 \gamk
\left< \left| \xiRkh \right|^2 \right> \left[ 1- 
\frac{2\Omega}{\Omega^{\prime} R} \,
\frac{\waz^2 - \gamk^2}{\waz^2 + \gamk^2} \right] \; .
\end{equation}
We note that the transport of angular momentum, produced by modes
of the considered type for which $\nabla \cdot \hat{\mathbf{v}} = 0$ and
evaluated from the relevant quasilinear theory assuming the adiabatic
equation of state, is not accompanied by an increase of thermal energy.
This is an important feature that differentiates the effects of these
modes from those of a viscous diffusion coefficient.

Using the dispersion relation (\ref{MRIdispersion}) we can verify that 
$\Deff$ is positive.
To assess the order of magnitude of $\Deff$ we may
argue, as is usually done in plasma physics, that, at saturation,
$|\xiRt| \sim 1/\kzo \sim 1/\ko$.  Thus we have
\begin{equation}
\label{deffestimate}
\Deff \sim \frac{\gam}{\ko^2} \sim
\left( \frac{\gam}{\Omega} \right) \frac{\va^2}{\Omega} \sim
\left( \frac{\gam}{\Omega} \right) c_s \frac{H}{\beta_s}  \; .
\end{equation}
Since the existence of these modes requires that $\beta_s \gg 1$ we
may conclude that this kind of mode packet should not produce a rate
of transport of angular momentum that is comparable with the one
represented by the Shakura-Sunyaev coefficient \citep{ss}
\begin{equation}
\label{dss}
\mathcal{D}_{ss} \simeq \alpha_{ss} c_s H \; .
\end{equation}
This is frequently used with ad hoc values of the numerical coefficient
$\alpha_{ss}$ that exceed $10^{-2}$.

\begin{figure}
\epsscale{0.5}
\plotone{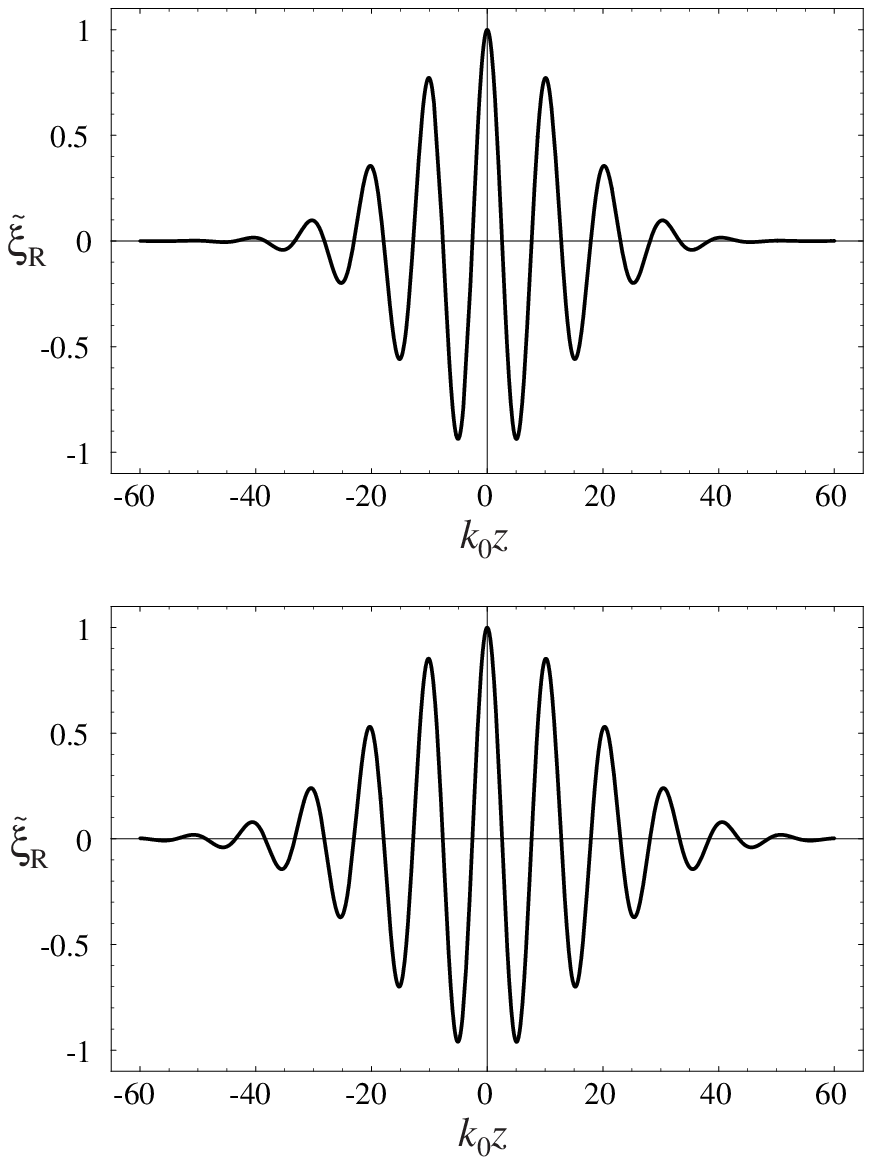}
\caption{
Spreading of non-normal mode packets.  An initially contained mode packet
formed from a range of $k_z$ centered around the peak growth rate of the
MRI will spread over time due to the dispersion relation.  The top panel
gives an initial packet with a width $\Delta = 10 \ko^{-1}$ at $t=0$,
and the bottom panel shows the same packet at $\gam t = 25$.  Here
$\ko = \sqrt{3} \Omega / \vaz$, and the width $\Delta$ is chosen by the
criterion that the mode packet undergoes minimal deformation as a function
of time while having the maximum growth rate.  The length is
given in units of $\ko^{-1}$, so one can see that the mode packet is
unacceptably spread even at $t=0$ for $\ko \Ho < 30$ (corresponding to
$\beta_s \sim 1000$), and even higher values of $\beta_s$ are necessary
to retain containment through many orbit times.
\label{fig:packet}}
\end{figure}

It is evident that the mode packets have the intrinsic limitations of not
being the normal modes of the system when considering their possible
excitation.  In particular we may consider their growth to be limited
to the case where
\begin{displaymath}
\Delta^2 - \frac{1}{2} \frac{\partial^2 \gam}{\partial k_z^2} t <
\frac{\Ho^2}{\aH^2}
\end{displaymath}
where $\aH$ is a reasonable numerical coefficient such as 3.  A numerical
derivation of the mode packet (\ref{modepacket}) deriving
$\partial^2 \gam / \partial k_z^2$ from the dispersion relation
(\ref{MRIdispersion}) is given in Fig.\ \ref{fig:packet}.  We note that
the packet spreads rather slowly as a function of time, as
$\Delta_{\mathrm{eff}}=[\Delta^2-(\partial^2 \gam / \partial k_z^2) t]^{1/2}$.
On the other hand the initial spread $\Delta$ has to be relatively broad
as the dispersion relation (\ref{MRIdispersion}) limits the range of values
of $k_z$ for which Eq.\ (\ref{disp_expansion}) is valid, as indicated in
Fig.\ \ref{fig:MRIdispersion}.  The consequence
of this is that the packet spread can reach the region where the Alfv\'{e}n
velocity varies significantly relative to its central value even at the
outset.

\placefigure{fig:packet}

When dealing with normal modes that are contained within the disk, given
the discrete spectrum of both $\kR$ and $\gam$ that characterize them and
that the vertical eigenfunctions are of the ballooning type, the standard
quasilinear theory cannot be applied to arrive at an estimate of
$\Deff$.  We may in fact use the expression
(\ref{deffestimate}) which can be arrived at by standard qualitative
considerations, and conclude that the relevant values of
$\Deff$ would fall well below those estimated
from Eq.\ (\ref{dss}).  On the other hand, the greatest difficulty
that the ballooning modes have is that on envisioning a process by which
they can be localized radially.

\acknowledgements

It is a pleasure to thank P.S. Coppi, with whom the present work was
started, for his continuing interest, and L.E. Sugiyama for comments.
This work was sponsored in part by the U.S. Department of Energy.

\appendix

%

\section{Higher Branch Solutions}

We note that additional branches of solutions to Eq.\ (\ref{diffeq}) also
exist at higher $\Lambda$ at exact intervals of 4, but with lower 
corresponding values of $\gamb^2$.  For completeness, the solutions for
branch $m \ge 0$ are
given by:
\begin{displaymath}
\tilde{Y}_{\mathrm{even}}(\bar{z}) = \sum_{i=0}^m a_{m,i}(\Lambda)\;
{}_1F_1 \left( i-m+\frac{\Lambda -3}{4} , \frac{1}{2} , -\frac{\bar{z}^2}{2} 
\right)
\end{displaymath}
\begin{displaymath}
\tilde{Y}_{\mathrm{odd}}(\bar{z}) = \bar{z}\;\sum_{i=0}^m b_{m,i}(\Lambda)\;
{}_1F_1 \left( i-m+\frac{\Lambda -1}{4} , \frac{3}{2} , -\frac{\bar{z}^2}{2} 
\right)
\end{displaymath}
where the coefficient functions are given recursively by:
\begin{displaymath}
a_{m,i}(\Lambda) = \frac{(i-m-1)(\Lambda+4i-4m-7)}{i(\Lambda+2i-4m-2)}\;
a_{m,i-1}(\Lambda)
\end{displaymath}
\begin{displaymath}
b_{m,i}(\Lambda) = \frac{(i-m-1)(\Lambda+4i-4m-5)}{i(\Lambda+2i-4m-2)}\;
b_{m,i-1}(\Lambda)
\end{displaymath}
with the base case given by normalization: $a_{m,0}(\Lambda) =
b_{m,0}(\Lambda) = 1$.  The dispersion relation for the higher
branches is
\begin{displaymath}
\gamb^2 = \frac{(\Lambda-4m-3)(\Lambda-4m-1)}{4}
\end{displaymath}
with $\Lambda > 3+4m$.


\clearpage

\clearpage

\clearpage

\clearpage

\clearpage

\clearpage

\clearpage

\clearpage

\begin{deluxetable}{ccl}
\tablewidth{0pt}
\tablecaption{Normal Mode Solutions
\label{modetable}}
\tablehead{
\colhead{$\Lambda$} & \colhead{$\gamb^2$} & \colhead{$\tilde{Y}(\bar{z})$}}
\startdata
5 & 2 & $e^{-\bar{z}^2/2}$ \\
7 & 6 & $\bar{z}\,e^{-\bar{z}^2/2}$ \\
9 & 12 & $(1-\bar{z}^2)\,e^{-\bar{z}^2/2}$ \\
9 & 2 & $(1+\frac{2}{3}\bar{z}^2)\,e^{-\bar{z}^2/2}$ \\
11 & 20 & $(\bar{z}-\frac{1}{3}\bar{z}^3)\,e^{-\bar{z}^2/2}$ \\
11 & 6 & $(\bar{z}+2\bar{z}^3)\,e^{-\bar{z}^2/2}$ \\
13 & 30 & $(1-2\bar{z}^2+\frac{1}{3}\bar{z}^4)\,e^{-\bar{z}^2/2}$ \\
13 & 12 & $(1+\bar{z}^2-\frac{2}{3}\bar{z}^4)\,e^{-\bar{z}^2/2}$ \\
13 & 2 & $(1+\frac{4}{19}\bar{z}^2+\frac{4}{19}\bar{z}^4)\,e^{-\bar{z}^2/2}$ \\
\enddata
\end{deluxetable}

\end{document}